# Ultrasensitive $^3$He magnetometer for measurements of high magnetic fields

A. Nikiel[1,2], P. Blümler[1], W. Heil[1,a], M. Hehn[3], S. Karpuk[1], A. Maul[1], E. Otten[1], L. M. Schreiber[4], and M. Terekhov[4]

[1]Institute of Physics, Johannes Gutenberg-Universität Mainz, Staudingerweg 7, D-55128 Mainz, Germany

[2]Helmholtz-Institute Mainz, Johann-Joachim-Becherweg 36, D-55128 Mainz, Germany

[3]MPI for Polymer Research, Ackermannweg 10, D-55128 Mainz, Germany

[4]University Medical Center Mainz, Langenbeckstrasse 1, D-55131 Mainz, Germany

**Abstract:**

We describe a $^3$He magnetometer capable to measure high magnetic fields ($B > 0.1$ Tesla) with a relative accuracy of better than $10^{-12}$. Our approach is based on the measurement of the free induction decay of gaseous, nuclear spin polarized $^3$He following a resonant radio frequency pulse excitation. The measurement sensitivity can be attributed to the long coherent spin precession time $T_2^*$ being of order minutes which is achieved for spherical sample cells in the regime of "motional narrowing" where the disturbing influence of field inhomogeneities is strongly suppressed. The $^3$He gas is spin polarized in-situ using a new, non-standard variant of the metastability exchange optical pumping. We show that miniaturization helps to increase $T_2^*$ further and that the measurement sensitivity is not significantly affected by temporal field fluctuations of order $10^{-4}$.

## 1. Introduction

Ultra-sensitive measurements and monitoring of high magnetic fields ($B > 0.1$ T) are of great interest for different fields of physics and applied research, ranging from accelerator science (e.g. MAP, Muon Accelerator Program [1,2]), to mass spectroscopy [3] and practical applications like shimming procedures for permanent and superconducting magnets [4] or gradient monitoring during MRI experiments [5].

[a] corresponding author : wheil@uni-mainz.de

Magnetic field strength is measured using a variety of different technologies. Each technique has unique properties which determine its suitability for particular applications. These applications can range from simply sensing the presence or change in the field to the precise measurements of a magnetic field's scalar and vector properties. An exhaustive fundamental description for sensing magnetic fields can be found in Ref.[6] .

The most straightforward way to verify that the required field strength, shape and desired homogeneity as well as its temporal stability, have been achieved is still a direct magnetic measurement via the method of nuclear magnetic resonance (NMR) [7]. Due to its capability of determining absolute field values, it is also commonly used for the calibration of magnetic sensors which are based on other physical principles such as, e.g., Hall plates.

The basic principle of NMR is to polarize the magnetic moments $\mu_I$ of nuclei along the axis of the respective magnetic field (z-axis), and then to tip them synchronously out of axis towards the transverse x-y plane by applying a short, resonant radio frequency (rf) pulse. Subsequently, the free, coherent precession of the nuclear magnetic moments around the field direction with the Larmor frequency

$$f_L = (\gamma/2\pi) \cdot B \tag{1}$$

is detected by means of an induction coil (pick-up coil). The proportionality constant $\gamma$ is called the gyromagnetic ratio and is a property of the respective nucleus. The loss of coherence in a process called Free Induction decay (FID) is usually characterized by an exponential decay with a time constant $T_2^*$. This time constant $T_2^*$ can range from a few ms to several seconds, depending on the nature of the sample and specific factors such as the presence of field gradients. The measurement time is thus limited to the order of $T_2^*$. The high resolution ($\Delta f_L / f_L$) of NMR magnetometers arises from the sharp resonance lines ($\Delta f_L = 1/(\pi \cdot T_2^*)$) observed in many liquid or gaseous samples containing nuclei with nonzero spin. Today, NMR provides the most reliable standard for the measurement of a homogeneous magnetic field commonly achieving accuracy of 0.1 ppm and better in controlled conditions [2,8,9].

New precision experiments currently in progress, such as the measurement of atomic masses by cyclotron resonance with a relative uncertainty of $\delta M/M \leq 10^{-11}$ demand the knowledge of the magnitude of magnetic fields with a similar accuracy [10]. This requires a basic improvement of measurement technology for the control and regulation of the magnetic field hitherto not been possible. Although such Penning trap experiments often are conducted at cryogenic temperatures ($T = 4K$) in relatively tight and complex structures, this should not principally affect the applicability of an NMR-magnetometer.

The key to a significantly increased sensitivity in NMR-magnetometry is the achievement of long $T_2^*$-times of coherent spin precession. The relatively short $T_2^*$ of liquid - and even more of solid samples is mainly due to mutual dipolar couplings [11] and can be greatly reduced when using gaseous samples. In doing so, the lower density of spins in a gas must be compensated by hyperpolarization, i.e., spin polarization that by far exceeds the thermal Boltzmann polarization, to give a useful NMR signal. [1]

The possible achievable accuracy of the frequency and thus magnetic field measurement can be estimated in the following solely statistical way: Assuming the noise is Gaussian distributed, the Cramer-Rao Lower Bound (CRLB) [12] sets the lower limit on the variance $\sigma_f^2$ for the frequency estimation of an exponentially damped sinusoidal signal given by

$$\sigma_f^2 \geq \frac{12}{(2\pi)^2 \cdot SNR^2 \cdot f_{BW} \cdot T^3} \cdot C(T, T_2^*), \qquad (2)$$

where $f_{BW}$ is the bandwidth of the acquisition of duration $T$. $SNR$ is the signal-to-noise ratio and $C(T, T_2^*)$ describes the effect of exponential damping of the signal amplitude with $T_2^*$. For observation times $T \leq T_2^*$, $C(T, T_2^*)$ is of order one [13]. The sensitivity $\delta B$ on the respective magnetic field $B$ seen by the sample spins is derived from Eq. 2 using Eq. 1. It increases with the observation time, $T$, according to

$$\delta B \geq \frac{\sqrt{12} \cdot C(T, T_2^*)}{\gamma \cdot SNR \cdot \sqrt{f_{BW}} \cdot T^{3/2}}. \qquad (3)$$

Due to their 2 to 3 orders of magnitude higher gyromagnetic ratio γ, Eq. 3 suggests the use of magnetometers based on the spin precession of electrons rather than on the spin precession of nuclei. Usually, the relaxation time of electron spins is short (~ ms), while nuclei, such as

---

[1] Compared to thermally polarized water ($^1H_2O$) with $P_H = 3.4 \cdot 10^{-6} \cdot B[T]$, the same NMR signal strength is obtained for gaseous HP spin samples ($P \approx 1$) at pressures of order *mbar*.

$^3$He, display a much longer spin-relaxation time. In practice, the product of $\gamma_{He} \cdot T^{3/2}$ exceeds that of electron spin based magnetometers for $T > 0.1$ s, already. Moreover, at high magnetic fields, where the frequencies of electron spin magnetometers are in the GHz range, elaborated microwave techniques are necessary for excitation and detection. With the expected $T_2^*$-times of ~ 1min for gaseous $^3$He continuous field monitoring over at least 3 characteristic time constants is possible. Taking an *SNR* of $SNR = 1000:1$ in a bandwidth of 1 Hz, $\gamma_{He} = 2\pi \times 32.4340921(97)$ [MHz/T] [14], and $T = T_2^* \approx 1$min, a lower limit of $\delta B \approx 40\,fT$ corresponding to a relative precision in magnetic field measurement of

$$\left(\frac{\delta B}{B}\right)_{\text{CRLB}} \approx 4 \cdot 10^{-14} / B[\text{T}] \tag{4}$$

is expected from Eq. 3. In the following sections we discuss how far this CRLB limit in measurement sensitivity is met under realistic conditions, in particular in situations where the magnetic field and with it the signal frequency may vary within the measurement time *T* by amounts $\delta f >> (\sigma_f)_{CRLB}$. The presented approach is similar to the pioneering work of Flowers et al. [14]. First and foremost, however, we focus on relative measurements of the temporal stability of the magnetic field rather than its absolute value.

## 2. Methodology

### 2.1 Concept of long spin relaxation in gaseous polarized $^3$He sample

$T_2^*$ in gases is strongly influenced by diffusion in inhomogeneous magnetic fields. Due to a low pressure of the $^3$He sample ($p \approx \mathcal{O}(\text{mbar})$) one can reach a high diffusion constant (*D*). In the so-called "motional narrowing" regime, where the gas atoms diffuse throughout the sample cell (spherical cell of radius *R*) in a relative short time $T_D \approx R^2/D << 1/(\gamma \Delta B)$, the disturbing influence of the field inhomogeneity ($\Delta B \approx R \cdot |\vec{\nabla} B|$) on the NMR coherence time $T_2^*$ is strongly suppressed. This was calculated for spherical cells by Cates et al. [15]. Following their notation, the general expression for the transverse relaxation rate $1/T_2^*$ is given by

$$\frac{1}{T_2^*} = \frac{1}{T_1} + \frac{1}{T_{2,field}}$$

$$= \frac{1}{T_1} + \frac{8R^4\gamma^2|\vec{\nabla}B_z|^2}{175 \cdot D} + D\frac{|\vec{\nabla}B_x|^2 + |\vec{\nabla}B_y|^2}{B^2} \times \sum_n \frac{1}{|x_{1n}^2 - 2| \cdot \left(1 + x_{1n}^4 (\gamma B R^2/D)^{-2}\right)}, \quad (5)$$

where $x_{1n}$ ($n = 1,2,3,...$) are the zeros of the derivative of the spherical Bessel function and $T_1$ is the longitudinal relaxation time which typically exceeds $T_2^*$ by far, being ~ 1h in small Pyrex and/or quartz cells. Hence, for our application only the latter, field dependent terms are relevant. At gas pressures $p$ around 1 mbar and high magnetic fields ($B > 0.1$ T), this formula reduces to

$$\frac{1}{T_2^*} = \frac{8R^4\gamma^2|\vec{\nabla}B_z|^2}{175 D_{0,^3He} \cdot \sqrt{T/273}} \cdot p, \quad (6)$$

where we have used $D = D_{0,^3He} \cdot (p_0/p) \cdot (T/273\,\text{K})^{1/2}$ [cm$^2$/s] with $D_{0,^3He} = 1880$ [cm$^2$/s] at $p_0 = 1$ mbar [16] for the self diffusion coefficient of $^3$He. That is, the duration of the FID-signal in addition to material constants depends on the radius $R$ of the sample cell, field gradient $|\vec{\nabla}B_z|$, temperature $T$, and gas pressure $p$.

Taking $|\vec{\nabla}B_z| \approx 10^{-7}$ T/cm and $R = 1$ cm, the expected $T_2^*$ is of order of minutes at $p = 1$ mbar and at room temperature. The $R^4$ dependence of $1/T_2^*$ suggests to miniaturize the sample cell, i.e., $R \approx \mathcal{O}(\text{mm})$. This has two positive side effects: i) Reducing $R$, will allow to operate the sensitive magnetometer at quite stronger magnetic field gradients, and ii) miniaturization will lead to an increased sensitivity [2]. However, in going to smaller cell sizes, care must be taken to disturbing influences of the magnetic susceptibility which is a well-known fact in high resolution NMR spectroscopy [18]. Of particular importance here is the spherical symmetry of the cell (cf. Sect. 4.1).

---

[2] According to Eq. 3, the sensitivity essentially scales $\propto SNR \times (T_2^*)^{3/2}$. To first order the drop in *SNR* with *SNR* ~ $R^2$ (Eq.10 of Ref.[17]) can be compensated by higher gas pressures and/or the use of detection coils with improved filling factors (cf. Sect. 3.2)

## 2.2   Polarization of gaseous $^3$He sample

Optical pumping transfers polarization from photons to atoms by resonant absorption of circularly polarized light. In metastability exchange optical pumping (MEOP) of $^3$He, a small fraction of the helium atoms (~ppm) are excited by a weak gas discharge into the metastable $^3S_1$-state, where they can absorb resonant circularly polarized laser light at 1083 nm, e.g., via the optical transition from $2^3S_1(F = 3/2)$ in $2^3P_0(F = 1/2)$ and thus be optically pumped [19]. Because of hyperfine coupling, the electronic polarization becomes a nuclear polarization. Finally, the metastability exchange collisions between metastable $^3$He atoms with $^3$He atoms in their ground-state ($1^1S_0$) transfer nuclear polarization to the ground-state.

The hyperfine interaction plays a crucial role in MEOP, providing both the physical mechanism for the polarization transfer during optical pumping from the atomic electrons to the $^3$He nuclei and causing nuclear polarization losses by back transfers of nuclear orientation to electronic orientations of exited states with $L \neq 0$. The high magnetic field decoupling minimizes the latter effect, counterbalancing the reduction of population of metastable atoms at higher gas pressures. That's why the MEOP technique can be extended to elevated gas pressures at high magnetic fields. For our application, it is more appropriate to work at low pressures $\mathcal{O}$(mbar), where one essentially benefits from the long spin coherence times $T_2^* \propto 1/p$ (cf. Eq. 6) and only secondarily on the *SNR* which increases with pressure. A detailed microscopic model for the MEOP at arbitrary magnetic field was developed [20], and systematically studied in fields up to 4.7 T, where the hyperfine interaction in the $2^3S_1$-state of $^3$He is still strong enough to allow hyperpolarization of $^3$He. However, when the Zeeman energy exceeds the fine- and hyperfine structure scales, the angular momentum structures of the $2^3S$ and $2^3P$ levels and the 1083 nm transition are deeply modified. Thus, one has to tune the pump laser to one of the strong absorption lines, called $f_2^\pm$ and $f_4^\pm$, respectively [21] with the superscripts ± standing for $\sigma^\pm$ light polarization that yield the highest polarization for magnetic fields $B > 1$ T.

While this discussion holds for room temperature, the situation changes again when decreasing the temperature. Under such conditions, the rate of metastability exchange collisions is drastically reduced, e.g., by a factor of about 30 by dropping the temperature from 300 K to ≈ 4 K [22,23]. That's why MEOP becomes inefficient and slow. We determined build-up times for the nuclear polarization of $^3$He in our small samples (see Fig. 1 below) at 300 K and 4.7 T in the order of seconds which would increase to 1 to 2 minutes at around 4 K without any further optimization. The elevated build-up times at low temperatures, however, should not be a big impairment for the applicability of this magnetometer. MEOP at cryogenic temperatures has been demonstrated in [24,25], that allows optical pumping inside the cold bore tube of , e.g., a Penning trap magnet.

## 3.　　Experimental

### 3.1　　Setup for Metastability Exchange Optical Pumping (MEOP)

Two different shapes of Pyrex cells (cylindrical: length × diameter = 30 × 34 mm; 20 × 25 mm and spherical with 15, 20 and 30 mm diameters) were filled with $^3$He at 1 mbar pressure and sealed leaving an unavoidable small stem. Two electrodes were mounted on the outer surface of the cells to produce the plasma discharge in the gas. A broadband (~ 2 GHz) ytterbium-doped fiber laser at λ ≈ 1083 nm with 2 W maximum output power (Keopsys, France) was used for optical pumping of $^3$He in high magnetic fields. Its spectral linewidth is matched to the Doppler width (FWHM) of the helium gas at room temperature.

The light from the laser was transferred to the optical table inside the magnet by a single mode fiber, then collimated and circularly polarized before entering the $^3$He cell. After having passed the cell, the beam was blocked outside of the magnet. MEOP at high magnetic field was performed using the strong $f_2^-$ doublet of Zeeman components of the $2^3$S-$2^3$P resonance line split by 1.37 GHz at $B$ = 1.5 T [26].

The measurement protocol did not include the monitoring of the achieved steady state $^3$He nuclear polarization in absolut values, e.g., by means of a probe laser as was done in [27]. In our examinations relative measurements of the $^3$He nuclear polarization were performed at 1.5 T and 4.7 T using the measured FID signal amplitudes after preceding optical pumping processes with preset durations. For that the $^3$He sample cell was manually removed from the electrodes and located on top of a surface coil for NMR detection of the FID. The build-up curve (Fig. 1) shows that the nuclear polarization grows exponentially: $P(t)=P_\infty [1 - \exp(-t/t_b)]$ with $t_b$ as the characteristic build-up time toward its steady-state value $P_\infty$. Under typical operating conditions[3], the build-up times are around 1 s and 4 s at 1.5 T and 4.7 T, respectively. This shows that optical pumping of small spin samples is fast enough even at high magnetic field strength to allow for an almost continuous monitoring of the magnetic field with short interruptions given by the build-up time $t_b$.

---

[3] spherical Pyrex cells of 20 mm diameter filled with $^3$He at ~ 1 mbar, 2 W laser power for optical pumping, and strong gas discharge as defined in [27].

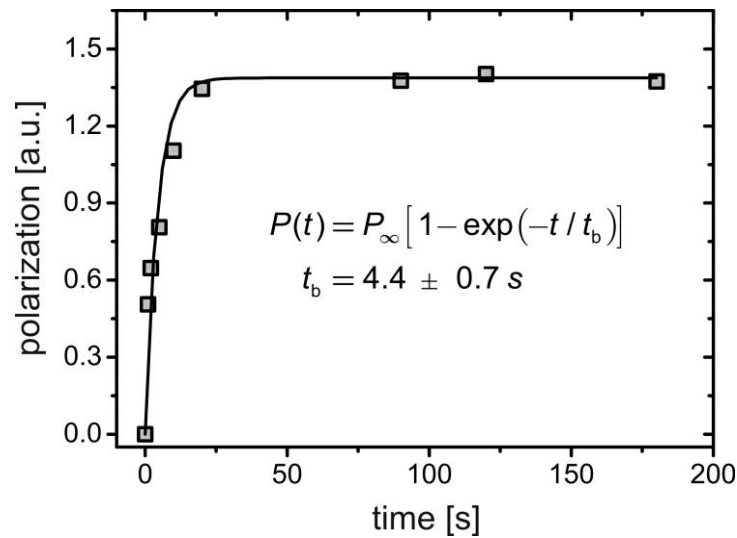

**Fig. 1.** Build-up of $^3$He nuclear polarization in the spherical Pyrex cells of 20 mm diameter with 2 W optical pumping laser power and strong discharge conditions in a magnetic field of 4.7 T. The measuring protocol was following: first the $^3$He gas was polarized for the time interval shown on the abscissa and afterwards an NMR-signal (in a.u. on the ordinate) was recorded which is proportional to the $^3$He-polarization. Solid curve: exponential fit to the data points (hollow squares).

### 3.2 NMR setup

All of the following experiments were performed on a 1.5 T whole body MRI-scanner (Magnetom Sonata, Siemens, Erlangen, Germany) in the Medical Center of the University Mainz. A dedicated experimental setup (pumping cell and the optical apparatus) used for high field MEOP was mounted on a non-magnetic table inside the 60 cm diameter bore of the magnet. The laser and all the electronic equipment were located a few meters away in a low magnetic field region.

The *SNR* of the NMR-signal together with its "life-time", $T_2^*$, are determining the achievable accuracy (cf. Eq. 2). At the same time, *SNR* and $T_2^*$ of the NMR-signal are independent parameters. To optimize the *SNR*, the excitation/detection coil should be as close as possible to the sample [17], while the direct vicinity of the NMR circuit board or other objects increases the magnetic field inhomogeneity in the coil/sample system and its local surroundings due to a susceptibility mismatch that, according to Eq. 6, may considerably shorten the $T_2^*$. Therefore, magnetic susceptibility related field inhomogeneities have to be eliminated by taking zero-susceptibility matched matter samples ($\chi_{mag} \approx 0$ ppm) as it is common practice in high resolution NMR spectroscopy [28].

The compromise we have chosen is to polarize the $^3$He-sample with attached discharge coils. After the polarization process was finished, the sample was manually moved to the detection circuit, which was located at the spot of highest magnetic field homogeneity (cf. Sect. 4.1). The $^3$He-sample was rested in a piece of Styrofoam tailored to its shape. In order to avoid disturbance of the field inside the samples by the direct vicinity of materials with different susceptibility we used a remote NMR coil for excitation and detection. This so-called surface coil was located about $d = 1$ cm below the samples[4], and has a single, quadratic loop of copper foil with a side length of 40 mm etched atop a glass-fiber board. The resonant circuit was tuned to $f_L = 48.6$ MHz, the Larmor frequency of $^3$He at 1.5 T and matched to the standard impedance of the connecting lines (50 Ohm). The resonant $Q$ of the coil and input circuit is $Q \approx 280$. As preamplifier a low noise GaAs FET (FSC60ML) was used together with a bandpass filter of 50 kHz. With this setup, we typically achieved SNR better than 1000:1 in a bandwidth of 1 Hz with flip angles according to 30°. Unfortunately, the used software (Siemens) restricted the acquisition to a maximum of 6.6 s.

Under the special conditions of this experiment, in particular the unusually long $T_2^*$-times, we have to consider the effect of radiation damping onto the transverse magnetization. According to Abragam [29], this effect is caused by the interaction between the sample's magnetization and the tuned rf resonator. It can be classically described as the action of a "counter" field produced by the current induced in the coil, which exerts a torque opposite to that of the transmit rf field to restore the net sample magnetization back towards the longitudinal direction. In the original analysis of Bloembergen and Pound [30], the temporal envelope of the radiation damped FID was derived as

$$S_{FID}^{RD} \propto \frac{2 \cdot \tan(\alpha/2)}{\exp(t/\tau_{RD}) + \tan^2(\alpha/2) \cdot \exp(-t/\tau_{RD})} \qquad (7)$$

where $\alpha$ is the flip angle and the damping time constant $\tau_{RD}$ in SI units is given by

$$\tau_{RD} = 2/(\eta \cdot Q \cdot \gamma \cdot \mu_0 \cdot M_0). \tag{8}$$

---

[4] at distances $d \geq 1$ cm, no shortening of the measured $T_2^*$-times could be observed anymore.

$M_0$ is the longitudinal magnetization and $\eta$ is the coil filling factor. The coil filling factor $\eta$ is a parameter that has different definitions and several practical means of its measurement are in use. Following Hill and Richards [31], we define the filling factor as the ratio of rf magnetic energy of the sample and the total stored magnetic energy in the surface coil, giving $\eta \approx 0.2\%$ for a spherical glass cell of $R = 1$ cm (setup as described above) filled with polarized $^3$He (we assume: $P_{He} = 0.2$) at $p = 1$ mbar. This results in $\tau_{RD} \approx 200$ s. Taking $\alpha \approx 30°$ for the flip angle, we derive for the envelope of the FID (including the exponential damping of the signal amplitude with $T_2^*$) an almost exponential decay:

$$S_{FID} \propto \exp(-t/T_{eff}) = \exp(-t/\tau_{RD}) \cdot \exp(-t/T_2^*), \tag{9}$$

i.e., the influence of radiation damping has to be taken into account for $T_2^* > 1$ min.

## 4. Results

### 4.1 $T_2^*$ dependence on sample shape and size

According to Eq. 6, $T_2^*$ depends on the size, $R$, and the field gradient, $|\vec{\nabla}B_z|$, inside the sample's container. In practice, the change in size is not completely decoupled from $|\vec{\nabla}B_z|$ and there are limits to a miniaturization of the container, as the $\sim 1/R^4$ dependence of $T_2^*$ may suggest.

First, exterior contributions of field inhomogeneities, e.g., from the superconducting magnet have to be considered. For that, the $^3$He magnetometer was moved both along the field axis and in radial direction around the center of the bore tube of the MR scanner. From the measured Larmor frequencies at different positions and by use of Eq. 1, the relevant field gradient $|\vec{\nabla}B_z|$ could be deduced to be $|\vec{\nabla}B_z|_{ext} \approx 3 \cdot 10^{-8}$ T/cm. Thus we expect a $T_2^*$ of about 1000 s under optimal conditions for a spherical cell of $R = 1$ cm filled with $^3$He at $p = 1$ mbar (Eq. 6).

It is well known that only infinitely long cylinders with their axis parallel to the external field and spheres will create no additional field gradients inside their inner volume [32].

Nevertheless, at first we used short cylindrical glass cells with optical windows for adjusting the optical polarization setup. Although the *SNR* of the acquired NMR-signals was very high, this shape was less favorable to generate very long $T_2^*$ values. This is because the edges at the front and the end of the cylinder create additional interior field gradients which destroy the spin coherence. We obtained $T_2^*$ values of only 3 s for a cylinder of 30 mm length and 34 mm diameter. Using an FEM-simulation (Comsol 3.4, Comsol Inc. USA) we estimated these gradients and calculated the $T_2^*$ values according to Eq. 6 which agreed well with the measurements.

To avoid such additional field gradients, spherical cells were used for further experiments. The polarization setup only needed small adaptations for this new sample geometry and similar *SNR*s were acquired. Although the increase in $T_2^*$ was significant using spheres instead of cylinders, it turned out that it was very difficult to obtain geometrically perfect sample containers. This is due to the fact that all the used glass containers were hand blown with limited precision and - more importantly – factured with a small glass capillary to evacuate and subsequently fill the container with $^3$He. Afterwards this capillary was sealed off with a torch leaving a small stem to the otherwise quite perfect spherical sample. While this remaining conical-shaped stem might be considered as a small deviation from spherical symmetry for larger spheres, its influence becomes increasingly noticeable when the dimensions of the spheres are reduced. Furthermore, the asymmetry introduced by this stem causes a strong dependence of the field gradients on its orientation relative to the external magnetic field. This situation is illustrated in Fig. 2.

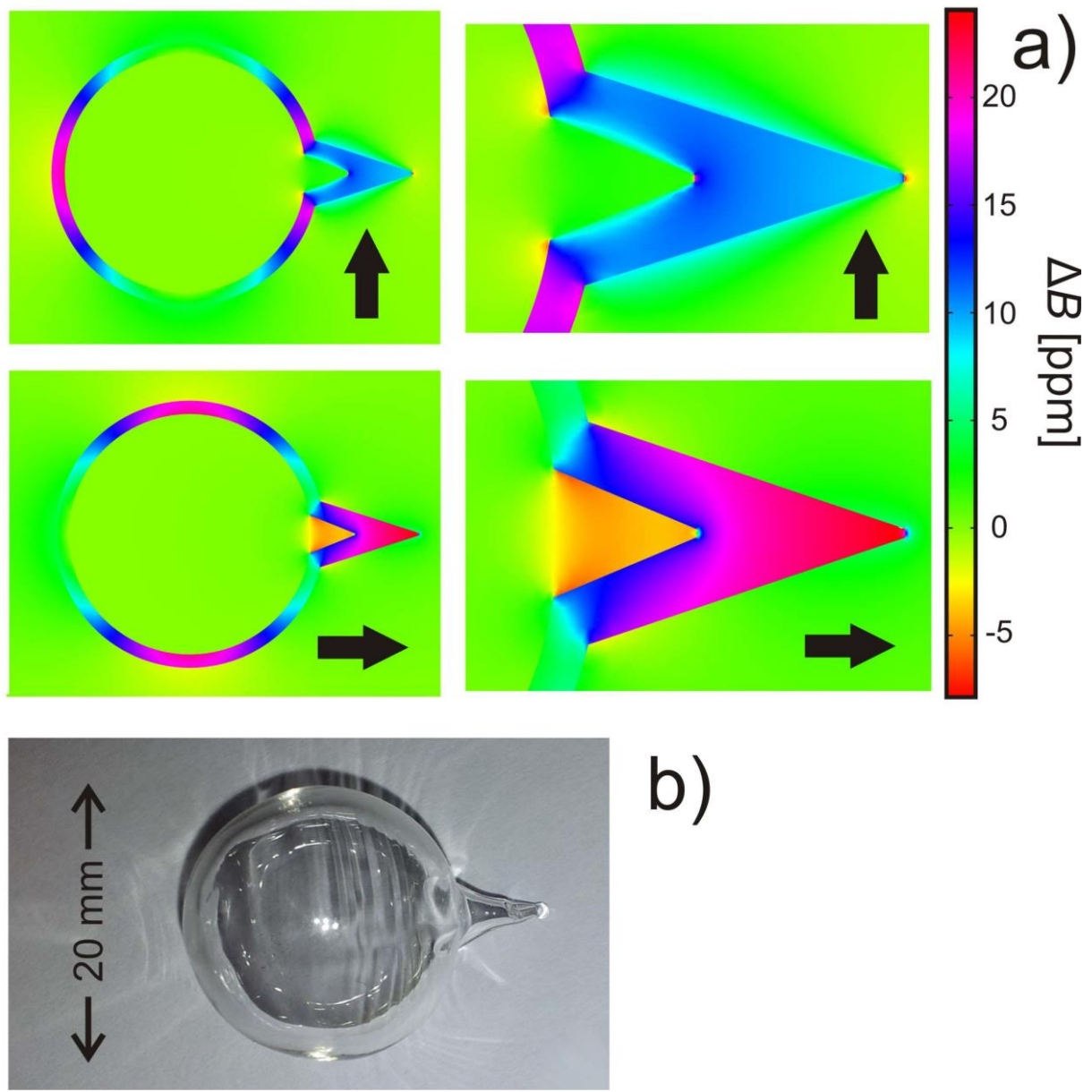

**Fig. 2**. **a)** Example of a 3D FEM-simulation of the magnetic flux inside a sphere (diameter 20 mm) with a small conical stem (inner diameter 3 mm, inner length 3.5 mm) to the right. Top row: Direction of flux perpendicular to stem. Bottom row: Direction of flux parallel to the stem. The black arrows indicate the flux direction. The right column shows a close up on the region of the stem. The colors mark flux differences in the ppm regime as indicated by the color-bar on the right. **b)** Photograph of a sealed off spherical glass cell with 20 cm diameter filled with $^3$He at $p = 1$ mbar.

It can be clearly seen that the flux homogeneity is worse when stem and flux are aligned. This observation was quantified by numerically calculating the gradient in flux direction and averaging its magnitude over the inner volume. From this $T_2^*$ can be calculated using Eq. 6. This was done for cell diameters of 15, 20 and 30 mm each with a stem of fixed size: 3 mm inner diameter and 3.5 mm inner length, which was a good approximation of the sample geometries used for the experiments (see Fig. 2b). Table 1 compares the experimental and simulated data.

**Table 1:** Measured $T_2^*$ of $^3$He ($p = 1$ mbar) at 1.5 T in comparison to the values from the numerical simulations for three spherical cells (1 mm wall thickness) with varying outer radius $R$ and the direction of the stem parallel and perpendicular to the direction of the external flux, $B$. $|\vec{\nabla} B_z|$ is the average gradient extracted from the simulations from which $T_2^*$ was calculated using Eq. 6. The influence of exterior contributions of field inhomogeneities to $|\vec{\nabla} B_z|$ was neglected as well as the relatively small influence of radiation damping on the measured $T_{eff} \approx T_2^*$ (cf. Eq. 9).

| cell radius $R$ [mm] | stem parallel to $B$ | | | stem perpendicular to $B$ | | |
|---|---|---|---|---|---|---|
| | $|\vec{\nabla} B_z|$ [T/cm] | calculated $T_2^*$ [s] | measured $T_2^*$ [s] | $|\vec{\nabla} B_z|$ [T/cm] | calculated $T_2^*$ [s] | measured $T_2^*$ [s] |
| 7.5 (6.5 int.) | $7.2 \cdot 10^{-7}$ | 11 | 8.65 | $3.85 \cdot 10^{-7}$ | 37 | 67 |
| 10 (9 int.) | $3.3 \cdot 10^{-7}$ | 14 | 10 | $1.8 \cdot 10^{-7}$ | 47 | 98 |
| 15 (14 int.) | $1.3 \cdot 10^{-7}$ | 15 | – | $1 \cdot 10^{-7}$ | 26 | 20 |

It can be seen that there is a fairly good agreement between simulation and experiment, especially when the stem is parallel to the external field and dominates the internal gradients. The numerical accuracy was tested by simulations of perfect spheres, where one expects no field gradients inside their inner volume. Gradient values in the order of $10^{-11}$ T/cm were found which can be regarded as a negligible gradient offset.

When inspecting Table 1 or Eq. 6 it is interesting to focus on the dependence of $T_2^*$ on the container size $R$, which is $T_2^* \propto R^{-4}$. The values in Table 1 do not directly show this trend because of the unproportionally stronger influence of the stem on the field inside smaller sample cells. Nevertheless, it can be expected that $T_2^*$ becomes extremely long in small cells of better sphericity.

## 4.2  Sensitivity to magnetic fields

Because our goal is the very precise monitoring of magnetic fields, we tested the magnetometer with a 20 mm sample cell by the following procedure: We placed the cell in the most homogeneous region of the whole body tomograph and added a simple Helmholtz coil (radius $r = 15$ cm, $n = 2$ turns) which provided a field shift of 12 pT/µA at its center. The Helmholtz coil was powered by a special, low noise current source inside the magnet which consisted of a standard battery in series with a choice of resistors delivering fixed currents of 10, 50, 100, 200, or 250 µA. The currents could be switched and inverted by galvanically isolated interruptors using optocouplers in order to avoid external noise currents. Figure 3 shows the beating of FID signals against a reference signal $f_R$ from the local frequency standard at beat frequencies

$$f_b = f_L - f_R \tag{10}$$

of roughly 0.5 Hz (left) and 0.1 Hz (right), respectively. They have been recorded for the maximum available acquisition time of 6.6 s. The measured $T_2^*$-time is $T_2^* \approx 70$ s. By help of the Helmholtz coil, the pair of FID signals in Fig. 3a has been split by applying a field of $|\Delta B_a^{set}| = 2.4$ nT ($i = 200$ µA) or $|\Delta B_b^{set}| = 0.6$ nT ($i = 50$ µA) as in Fig. 3b. The beat frequencies of the pair can be roughly evaluated from the swept time intervals for, e.g., 3 complete cycles in a) and 0.5 cycles in b). Estimating an error of 0.05 s for this reading we evaluate for pair a) a frequency difference of $|\Delta f_a| = 3 \cdot (1/(5.60(5)\text{s}) - 1/(6.45(5)\text{s})) = 0.071(6)$ Hz $\Rightarrow |\Delta B_a| = 2.20(18)$ nT. For pair b) we get $|\Delta B_b| = 0.62(4)$ nT. Within the error bars these results correspond to the set $\Delta B$-values. Note that the faster beating in case a) yields about a 4 times larger uncertainty (180 pT) as compared to the slower beating in case b) (40 pT). In the latter case this rough and ready evaluation reaches already a sensitivity limit for detecting relative field shifts as a small as $\Delta B / B \approx 3 \times 10^{-11}$ at $B = 1.5$ T.

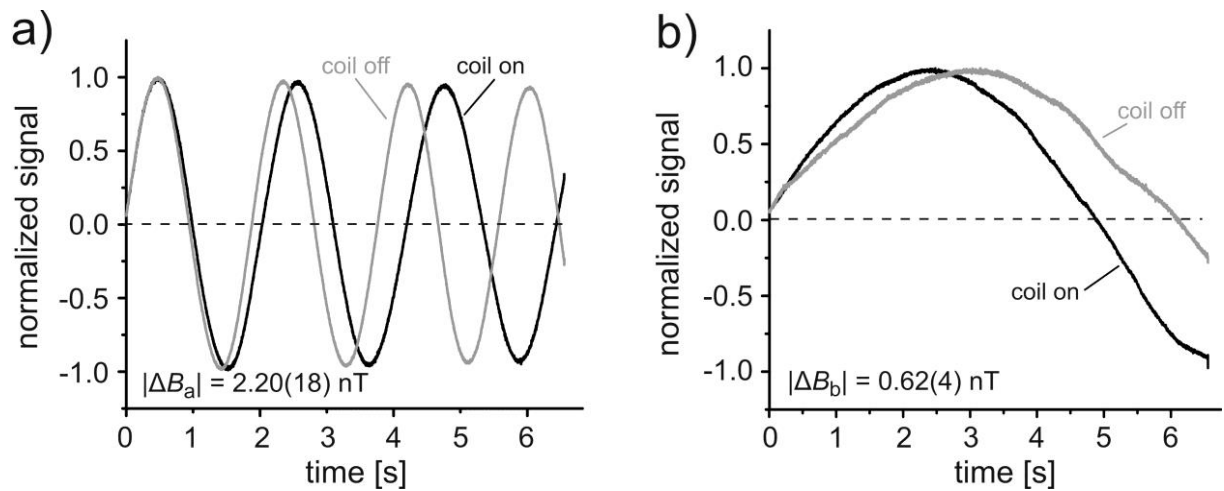

**Fig. 3**. Measured FID (normalized signal; sampling rate: $r_{s,0} = 620$ Hz) of the beat frequency $f_b$ with and without preset magnetic field shifts of a) $\left|\Delta B_a^{set}\right| = 2.4$ nT and b) $\left|\Delta B_b^{set}\right| = 0.6$ nT added to the $B = 1.5$ T field of the MR scanner. The characteristic time constant of the FID could be determined to be $T_2^* \approx 70\ s$. For technical reasons only 6.6 s of the FID could be recorded.

On closer inspection, one should not be surprised to see residual field fluctuations ("visible wiggles" in Fig. 3b) during monitoring of the FID which shift the measured $|\Delta B|$ by 50($\pm$40) pT off the set value. These are mainly caused by environmental field instabilities due to movement of elevators etc. and a clear "day- night" effect could be observed! Therefore, in a second approach, the currents were manually switched and inverted during the 6.6 s acquisition period, instead of extracting the preset field changes from FID measurements which were recorded separately.

Because the NMR signal is registered as a complex transient, the phase can be directly calculated (removing phase jumps by standard phase unwrapping algorithms implemented in MATLAB, Mathworks USA). The free induction decay of $^3$He can be described as

$$S(t) = S_0 \cdot \exp\left(-i\phi(t)\right) \cdot \exp\left(-t/T_2^*\right). \tag{11}$$

Then the phase is simply calculated by

$$\phi(t) = \tan^{-1}\frac{\Re(S(t))}{\Im(S(t))} = \int \omega(t)\,dt = \int \gamma\, B(t)\,dt. \quad (12)$$

The slope or the first time derivative of $\phi(t)$ is then the magnetic field multiplied by the gyromagnetic ratio.

About $n = 10$ sequential FID experiments in series (flip angle 30°) could be performed for each current setting on the same hyperpolarized $^3$He sample before the signal became too weak ($S_n \approx S_0 \cdot \cos^{n-1}(\alpha)$). The results are shown in Fig. 4. While the data nicely agree with the expectations on $\Delta B$, it can also clearly be seen that the errors are rather big. This is mainly due to the still noticeable environmental fluctuations of the magnetic field as evidenced by Fig. 3b. From the linear phase fit

$$\Phi(t) = \phi_0 + (2\pi \cdot f_b) \cdot t \quad (13)$$

to the data (see Fig. 4a) we could extract both $f_b^{\Delta B}$ and $f_b^{\Delta B=0}$. In Fig. 4b the quoted error on the mean $\overline{\Delta B} = \frac{1}{n}\sum_{i=1}^{n} 2\pi \cdot \overline{\left| f_b^{\Delta B} - f_b^{\Delta B=0} \right|_i} / \gamma$ essentially reflects these temporal field drifts.

Therefore, the chosen method to extract preset field changes via the Helmholtz coil, is limited and doesn't reflect the true sensitivity of this magnetometer. A more elaborated model is needed to describe the sensitivity of a magnetometer in case the magnetic field is not constant in time.

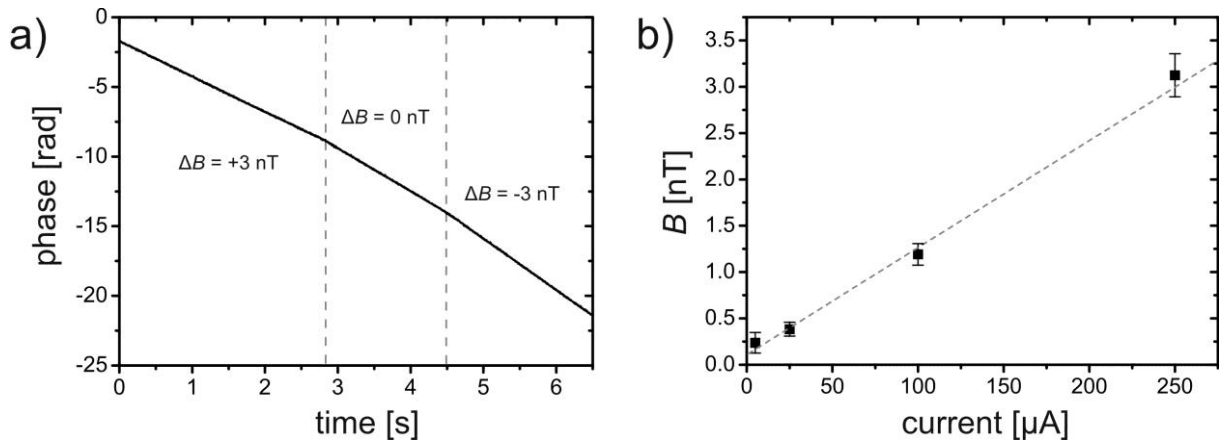

**Fig. 4.** a) Typical acquisition of the signal phase. During the 6.6 s acquisition period a current of about 250 µA was first switched on (ca. 0 - 3 s), then switched off (ca. 1.5 s) and finally inverted. Linear regression to these three sections yields the respective field change $\Delta B$ at the spot of the sample. This was repeated ten times for each current and the extracted fields (average) are plotted versus the current in b). The gray dashed line is a linear fit to the data.

### 4.3 Estimation of the achievable sensitivity

The use of CRLB for frequency estimation as introduced in Sect. 1 (Eq. 2) is based on $f$ = const., i.e., $B(t)$ = const. Now, the signal frequency from the source (in our case the beat frequency $f_b$) may vary within the measurement time $T$ by amounts $\delta\!f_b \gg \sigma_f^{CRLB}$. In this case, one may wish to measure the frequency averaged over the period $T$ represented by

$$\overline{f}_b = \frac{\Delta\phi}{2\pi \cdot T} \,. \tag{14}$$

$\Delta\phi$ is the accumulated phase as shown in Fig. 5 which can be expressed as

$$\Delta\phi = 2\pi \cdot m + \phi_F - \phi_I \,, \tag{15}$$

where $m$ is the number of phase jumps ($2\pi$) within $T$ and $\phi_{F,I}$ are the respective values of the phase determined at the beginning (I) and at the end (F) of the data train. The task now is to estimate the error on $\phi_{F,I}$ with maximum efficiency.

The two examples given in Fig. 5 show the acquired phase data from two FID runs called $A$ and $B$ (with Helmholtz-coil off: $\Delta B = 0$). At the sensitivity scale of *rad*, a linear functional dependence is the finding. Subtracting the function

$$g(t) = 2\pi \cdot \overline{f}_b^{A,B} \cdot t + \phi_I, \tag{16}$$

we obtain the corresponding phase residuals $\phi(t)$ which vary within $\approx$ 150 mrad. The time derivative of which gives the change in Larmor precession frequency or magnetic flux vs. time. From that a maximum field change of about 0.5 nT caused by environmental fluctuations of the magnetic field was estimated for these runs.

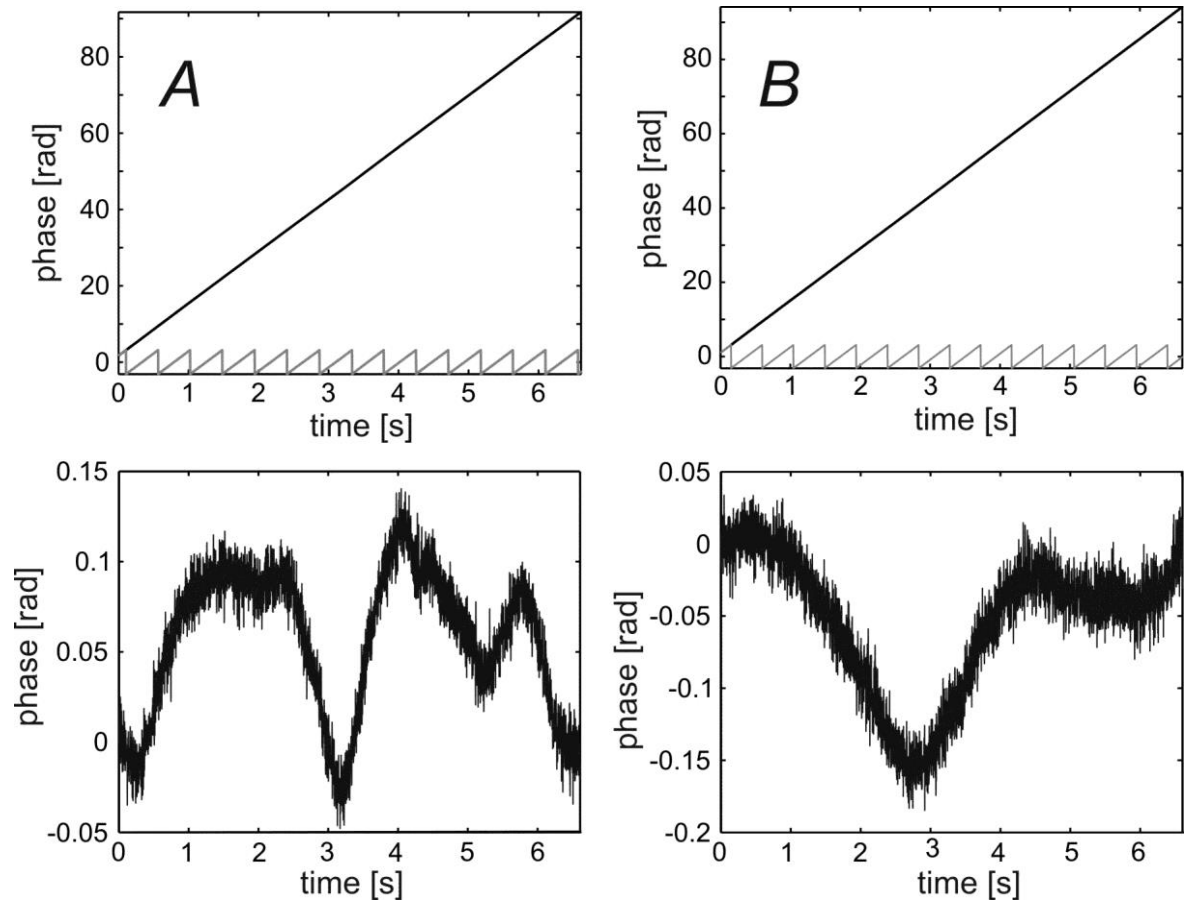

**Fig. 5.** Top: Acquired phase data from two FID runs (*A* and *B*) with Helmholtz-coil off, $\Delta B = 0$. During the acquisition time of $T = 6.6$ s, the accumulated phase ("unwrapped" data in black) amounts to $\Delta \phi \approx 90$ rad. The number *m* of phase jumps ($2\pi$) is extracted from the arctangent function that gives the wrapped phase variation (sawtooth-like structure in gray, here $m = 15$). Bottom: Phase residuals after subtraction of Eq. 16 indicate the environmental phase noise induced by field changes of $\leq 0.5$ nT.

These phase residuals are used to estimate the error on $\phi_{F,I}$. Thereby, the Allan Standard Deviation (ASD) method is applied for the analysis of signal noise and drift. The ASD [33] is the most convenient measure to study the temporal characteristics of frequency fluctuations and to identify the power-law model for the phase-noise spectrum under study. The ASD of the phase-residuals, $\phi(t)$ of the two FID runs (*A*,*B*) shown in Fig. 5, is calculated according to

$$\sigma_{ASD}(\tau) = \sqrt{\frac{1}{2(N-1)} \sum_{i=1}^{N-1} \left(\bar{\varphi}_{i+1}(\tau) - \bar{\varphi}_i(\tau)\right)^2} , \qquad (17)$$

where the total acquisition time $T$ is subdivided in $N$ smaller time intervals of the same length $\tau$, so that $N\tau = T$. For each such sub-dataset ($i = 1, 2,..., N$-1), the mean phase $\bar{\phi}_i(\tau) = \langle \phi_i(t) \rangle_\tau$ is determined. For white Gaussian noise −one essential requirement the derivation of CRLB is based on− $\sigma_{ASD}$ coincides with the classical standard deviation and we expect a $\sigma_{ASD} \sim \tau^{-1/2}$ dependence on the integration time $\tau$. This power-law is also found in our data for short integration times $\tau_m \leq 20$ ms ($\tau_m \leq 40$ ms) shown by the dashed (dotted) line in Fig. 6, whereas $\sigma_{ASD}$ increases again for $\tau > \tau_m$ due to the temporal characteristics of external field fluctuations that are the dominant sources of non-statistical phase fluctuations.

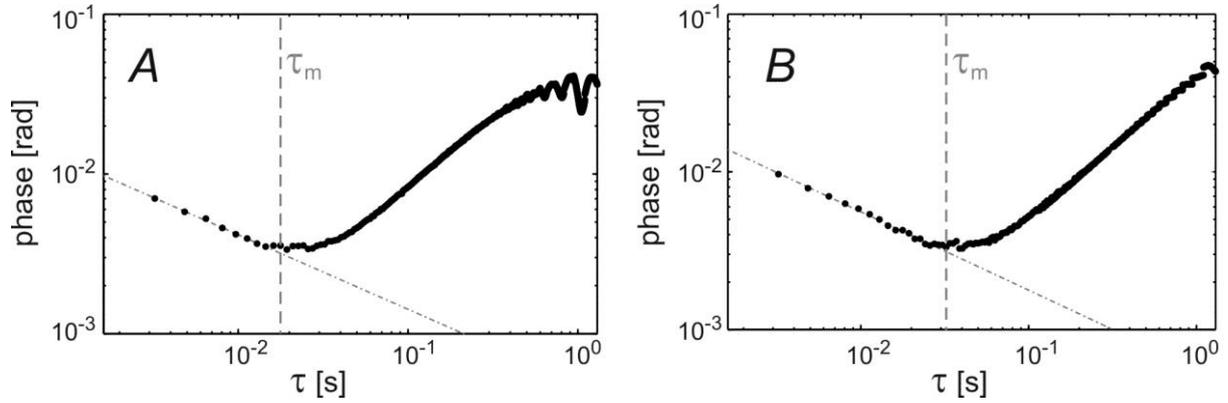

**Fig. 6.** Allen standard deviation (ASD) of the phase residuals measured for two FID-runs (*A*, *B*) after sequential NMR excitations of the same spin sample with flip-angle $\alpha \approx 30°$. The total observation time is $T = 6.6$ s. For integration times $\tau_m < 20$ ms ($\tau_m < 40$ ms), the residual phase noise decreases with $\tau^{-1/2}$ as indicated by an according fit (dashed lines). Beyond the respective $\tau_m$, the temporal characteristics of external field fluctuations causes the observed increase of non-statistical phase fluctuations. To fulfil the ASD statistics criteria [34], only data for integration times $\tau \leq 1$s are shown.

From the ASD-plot one can extrapolate the respective phase noise $\sigma_{A,B}$ at the sampling rate $r_{s,0} = 620$ Hz corresponding to $\tau = 1/r_{s,0} \approx 1.6$ ms : $\sigma_A \cong 9.8$ mrad and $\sigma_B \cong 13.8$ mrad. This phase noise defines the error on $\phi_{I,F}$. Using Eq(s). 13 and 14, the error on the average frequency $\bar{f}_b$ can be extracted to be $\sigma_{\bar{f}_b} = \sqrt{2} \cdot \sigma_{A,B} / (2\pi \cdot T) \cong 3.5 \times 10^{-4}$ Hz with $\sigma_{\phi_I} = \sigma_{\phi_F} = \sigma_{A,B}$. Finally, by use of Eq. 10, we obtain a first hint on the true sensitivity of this magnetometer: The accuracy with which the average Larmor frequency $\bar{f}_L$ and thus the average magnetic field $\bar{B} = 1.5$ T can be determined over the period of $T = 6.6$ s is given by $\delta B / \bar{B} = \delta f_L / \bar{f}_L \approx \sigma_{\bar{f}_b} / 48.6 [\text{MHz}] \approx 10^{-11}$. Here, we assumed that the frequency $f_R$ of the local oscillator (atomic clock in principle) is free of error.

A more refined analysis on the measurement sensitivity can be done if we determine the value $\phi_{I,F}$ from a group of data extending over $\Delta T$ at the beginning (I) and at the end (F) of the data train [35]. The ASD-plots in Fig. 6 show that the „zero-mean Gaussian noise" criterion CRLB is based on can be used for time intervals $\Delta T$ up to $\Delta T \approx 20$ ms and $\Delta T \approx 40$ ms, respectively. Within that time interval the phase $\phi_{I,F}$ can be set to be constant. According to [12], the phase error is estimated to be $\sigma_{\phi_{I,F}} = \sqrt{\sigma^2 / N_0}$ with $N_0 = \Delta T \cdot r_s$ and $\sigma$ being the phase noise at the sampling rate ($r_s$) limited bandwidth $f_{BW} = r_s / 2$ (see Fig. 6). From that the error on $\bar{f}_b$ can be derived to be

$$\sigma_{\bar{f}_b} = \frac{\sqrt{(\sigma_{\phi_I}^2 + \sigma_{\phi_F}^2)/N_0}}{2\pi \cdot T_m} \cong \frac{\sqrt{2} \cdot \sigma / \sqrt{\Delta T \cdot r_s}}{2\pi \cdot T} \tag{18}$$

with $T_m = T - \Delta T$ and taking $T >> \Delta T$.

The relative accuracy to which the average Larmor frequency and thus the average magnetic field can be determined over the total acquisition time $T$ is

$$\frac{\delta B}{\bar{B}} = \frac{\sigma_{\bar{f}_b}}{f_R + \bar{f}_b} \cong \frac{\sigma}{\pi \sqrt{2 \cdot r_s} \cdot T \cdot \sqrt{\Delta T} \cdot f_R} . \tag{19}$$

For instance, using values $T = 6.6$ s, $\Delta T = 40$ ms, $r_{s,0} = 620$ Hz, $\sigma_0 = 13.8$ mrad, and $f_R = 48.6$ MHz ($\bar{B} = 1.5$ T) this results in

$$\frac{\delta B}{\bar{B}} \cong 2 \cdot 10^{-12} . \tag{20}$$

The gain in sensitivity compared to the first estimate was achieved by the factor $\sqrt{N_0} \approx 5$, representing the increased statistical accuracy to estimate the error on $\phi_{F,I}$ with $N_0$ being the number of data points within the time interval $\Delta T$. It can be shown that Eq. 18 leads to the CRLB on the variance for the frequency estimation in the limit of a purely random distribution of the residual phase noise over the total acquisition time $T$, i.e., $\Delta T = T$.

Some practical considerations should be stated at this point: For long data acquistion times $T \approx T_2^*$ it may be advantageous to perform a separate ASD analysis over a given time slot ($\Delta T_{ASD} = \mathcal{O}(s)$) at the beginning and at the end of the FID run. The error on $\phi_{I,F}$ can then be determined independently from suitably chosen time intervals $\Delta T_{I,F}$. The latter ones strongly depend on the actual field fluctuations (non-statistical noise), but are not affected if field fluctuations happen during the relatively long data train in between the preset time slots used for the ASD analysis.

## 4.4    Dynamic range of field monitoring

The data presented were measured at an almost constant magnetic field of $\overline{B} = 1.5$ T with disturbing influences of the environmental fields being of order nTesla. The legitimate question may arise which dynamic range of field fluctuations can be covered by this magnetometer without significantly affecting its sensitivity. The precise measurement of the accumutated phase as discussed in the previous section presupposes that during data acquisition deviations $\Delta f_L$ from the mean value $\overline{f}_L \cong f_R$ of the Larmor frequency which are equivalent to the changes $\Delta f_b$ of the beat frequency after rf mixing with a carrier frequency $f_R$, have to fulfil the Nyquist-Shannon sampling theorem

$$\Delta f_b \leq r_s / 2 \equiv f_{Ny}. \tag{21}$$

With $\Delta f_b \leq 25$ mHz, the number which can be deduced from the phase residuals (time derivative) as shown in Fig. 5, the above requirement is met by far for the chosen sampling rate of $r_{s,0} = 620$ Hz. In principle, the sampling rate ($r_s$) can be noticeably increased in order to extend the dynamic range further. This, however, is associated by an increase of the phase noise (white noise) given by

$$\sigma = \sigma_0 \cdot \sqrt{\frac{r_s}{r_{s,0}}}, \tag{22}$$

where $\sigma_0 \approx 10$ mrad is the measured phase noise at $r_{s,0}$ as deduced from Fig. 5. Since the measurement precision is directly related to the error $\sigma_{\Delta\phi}$ in the accumulated phase, we take Eq. 15 as starting point and derive

$$\sigma_{\Delta\phi} = \sqrt{\sigma_I^2 + \sigma_F^2 + (2\pi \cdot \Delta m)^2} \ . \qquad (23)$$

Besides the phase errors $\sigma_{I,F}$, we also have to consider noise induced phase wraps $\Delta m$ which get more and more important the larger the dynamic frequency range that must be covered by the magnetometer. Already for $\Delta m = 1$, $\sigma_{\Delta\phi}$ is dominated by the latter effect since we have $\sigma_{I(F)} << 2\pi$.

The instantaneous phase signal (Eq. 12) can only take on values in a [-π, +π] range. This „wrapping" leads to discontinuities of 2π for each period of the complex signal. To recover the original phase, i.e., the accumulated phase (up to some global offset) requires an unwrapping algorithm which removes the modulus 2π ambiguities of the wrapped phase. It is the distinction between true or genuine phase wraps ($m$) and apparent or fake phase wraps ($\Delta m$) that have been caused by phase noise that make the practical phase unwrapping such a challenging task. A good review of different classes of algorithms is given by Robinson [36]. In our analysis we used the standard procedure embedded in MATLAB (ver. 7, Mathworks, USA) for phase unwrapping.

For practical reasons it is advantageous to express the phase noise $\sigma$ in terms of the *SNR* of the measured FID signal: $SNR = (\sqrt{2} \cdot \sigma)^{-1}$ [37]. By use of Eq(s). 21 and 22, the Nyquist frequency $f_{Ny}$ can then be written as

$$f_{Ny} = \frac{1}{2} \cdot \left( \frac{SNR_0}{SNR} \right)^2 \cdot r_{s,0} \qquad (24)$$

which gives, e.g., $f_{Ny} \cong 45$ kHz for an $SNR = 6$ using $SNR_0(r_{s,0}) \approx 72$. For different *SNR* we can simulate the number $\Delta m$ of fake wraps at beat frequencies $\Delta f_b = \kappa \cdot f_{Ny}$ with $0 \le \kappa \le 1$. This is shown in Fig. 7 (right) where $N = 10^7$ periods were analysed. If only a genuine phase wrap is the finding, the accumulated (unwrapped) phase $\Delta\phi = 2\pi \cdot m$ should reproduce $m = N$. Deviations $\Delta m = |m - N|$ are thus be attributed to fake wraps that have been produced by phase noise.

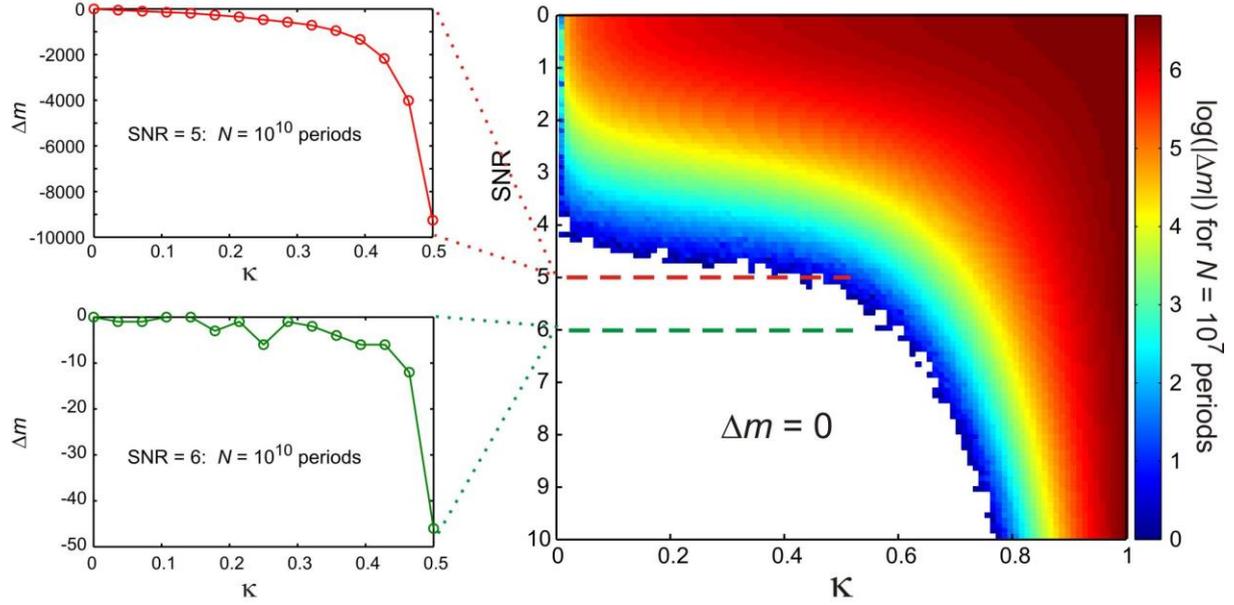

**Fig. 7.** Right: Contour plot (logarithmic color scale, i.e. $\log_{10}(|\Delta m|)$, no color for $\Delta m = 0$) of apparent phase wraps $\Delta m$ during $N = 10^7$ periods as a function of the *SNR* and the dynamic range, $\kappa$, of beat frequency fluctuations in units of the Nyquist frequency $f_{Ny}$. Left: Selected results for $\Delta m$ (linear scale) at an *SNR* of 5 and 6, respectively, after accumulation of $N = 10^{10}$ periods and a dynamic range to half the Nyquist frequency ($\kappa \leq 0.5$).

The logarithmic color scale of Fig. 7 (right) demonstrates the high error probability of phase unwrapping by going both to lower *SNR* and to beat frequencies $\Delta f_b$ approaching the Nyquist frequency. As an example, in Fig. 7 (left) are plotted the observed fake wraps $\Delta m$ versus $\kappa = \Delta f_b / f_{Ny}$ at an *SNR* of 5 and 6, respectively, each for a total of $N = 10^{10}$ periods. This result can be interpreted as a fidelity $F$, the degree to which $\Delta m' = 0$ is reproduced in phase unwrapping during the acquisition time $T$:

$$F = 1 - |\Delta m| \cdot \frac{\Delta f_b \cdot T}{10^{10}} \quad . \tag{25}$$

With the examples in Fig. 7 (left), one obtains, e.g., for $\Delta f_b = 0.45 \cdot f_{Ny}$ a fidelity of $F(SNR = 6) = 99.8\%$ ($|\Delta m| = 10$, $\Delta f_b \approx 20\,\text{kHz}$) and $F(SNR = 5) = 10\%$ ($|\Delta m| \approx 3000$, $\Delta f_b \approx 30$ kHz), respectively, during $T = 100$ s.

Therefore, an $SNR = 6$ enables monitoring of a dynamic frequency range of $|\Delta f_b| \leq 20\,\text{kHz}$ where the fidelity is practically 100%. This number determines the upper limit of allowed field fluctuations $\Delta B$ in the measurement interval $T$ ($\bar{B} = 1.5\,\text{T}$):

$$\frac{\Delta B}{\bar{B}} \leq \frac{|\Delta f_b|}{\bar{f}_L} \approx 4 \cdot 10^{-4}. \tag{26}$$

For $\Delta m = 0$, the measurement sensitivity is then solely determined by the phase noise $\sigma_{\Delta\phi} = \sqrt{\sigma_I^2 + \sigma_F^2} \approx \sqrt{2} \cdot \sigma$ which is $\sigma_{\Delta\phi} \approx 0.16$ for $SNR = 6$. As a result, we can derive the sensitivity to which the field $\bar{B}$ averaged over the period $T$ can be measured if we allow for relative field fluctuations of 0.04% (Eq. 26):

$$\frac{\delta B}{\bar{B}} \approx \frac{\sigma_{\Delta\phi}/(2\pi \cdot T)}{\bar{f}_L} \approx 5.5 \cdot 10^{-10}/\text{T}. \tag{27}$$

Already for an acquisition time of $T = 6.6\,\text{s}$ the relative measurement precision falls below $10^{-10}$. This accurarcy may be further improved if one determines $\sigma_{I,F}$ from a group of data extending over $\Delta T$ at the beginning and at the end of the data train as described in Sect. 4.3.

## 5. Conclusion and outlook

In this paper we have presented an ultra-sensitive $^3$He magnetometer to monitor magnetic fields of order Tesla to a relative precision of $\delta B/B < 10^{-12}$. The $^3$He gas is spin polarized in-situ using a new, non-standard variant of the metastability exchange optical pumping. Our approach is based on the NMR measurement of the free induction decay of nuclear spin polarized $^3$He after a resonant radio frequency pulse excitation. The use of spherical $^3$He sample cells embedded in an almost zero-susceptibility matched environment provides us with extraordinary long coherent spin precession times $T_2^*$ of more than one minute. Further miniaturization of the sample size ($R$) helps to be less sensitive to residual magnetic field gradients $|\vec{\nabla} B_z|$ since the transverse relaxation rate scales like $1/T_2^* \propto R^4 \cdot |\vec{\nabla} B_z|^2$ in the regime of motional narrowing, i.e., at gas pressure of order *mbar*. The sensitivity of this magnetometer which was tested inside the homogeneous field of an MRI scanner is not significantly affected if one allows magnetic field fluctuations of order $10^{-4}\,\text{T}$.

Its range of application can be extended to cryogenic temperatures, e.g., inside the cold bore tube of a Penning trap magnet. Almost all other NMR substances are solid at these temperatures in which the dipolar interaction between the nuclear spins leads to a dramatic decrease of $T_2^*$ (< 1 ms).

We are grateful to our glass blower R. Jera for preparing the spherical glass cells. This work was supported by the Deutsche Forschungsgemeinschaft (DFG) under the contract number HE2308/16-1 and by the Helmholtz-Institut Mainz, Section „Symmetry of Matter and Antimatter (MAM). Financial support from the Johannes Gutenberg-Universität Mainz in the framework of the "Inneruniversitäre Forschungsförderung" is greatfully acknowledged.